\documentclass[11pt]{article}
\usepackage{moriond,epsfig}

\def\be{\begin{equation}}
\def\ee{\end{equation}}
\def\bea{\begin{eqnarray}}
\def\eea{\end{eqnarray}}

\begin{document}
\vspace*{4cm}
\title{Recent results from the Milagro TeV gamma-ray observatory}

\author{P.~M.~Saz Parkinson for the Milagro Collaboration~\footnote{R.~Atkins,  
W.~Benbow, 
D.~Berley, 
E.~Blaufuss, 
D.~G.~Coyne, 
T.~DeYoung, 
B.~L.~Dingus, 
D.~E.~Dorfan, 
R.~W.~Ellsworth, 
L.~Fleysher,
R.~Fleysher,
M.~M.~Gonzalez,
J.~A.~Goodman,
T.~J.~Haines,
E.~Hays,
C.~M.~Hoffman,
L.~A.~Kelley,
C.~P.~Lansdell,
J.~T.~Linnemann,
J.~E.~McEnery,
A.~I.~Mincer,
M.~F.~Morales,
P.~Nemethy,
D.~Noyes,
J.~M.~Ryan,
F.~W.~Samuelson,
P.~M.~Saz Parkinson,
A.~Shoup,
G.~Sinnis,
A.~J.~Smith,
G.~W.~Sullivan,
D.~A.~Williams,
M.~E.~Wilson,
X.~W.~Xu
and 
G.~B.~Yodh}}

\address{Santa Cruz Institute for Particle Physics, 
University of California \\ 1156 High Street, 
Santa Cruz, CA 95064, USA}

\maketitle\abstracts{
Milagro is a gamma-ray observatory employing a water Cherenkov 
detector to observe extensive air showers produced by high-energy particles 
impacting in the Earth's atmosphere. We discuss the first detection of TeV gamma-rays 
from the Galactic plane and report the detection of an extended TeV 
source coincident with the EGRET source 3EG J0520+2556, and the  
observation of TeV emission from the Cygnus region of our Galaxy. 
We also summarize the status of our search for Very High Energy (VHE)
emission from satellite-triggered Gamma Ray Bursts (GRBs) and 
discuss plans for the next generation water Cherenkov detector.
}

\section{Introduction: The Milagro Observatory}

Milagro\cite{atkins01} is a TeV gamma-ray detector which uses the water Cherenkov 
technique to detect extensive air-showers produced by Very High Energy 
(VHE, $>$100 GeV) gamma rays as they interact with the Earth's atmosphere. 
Milagro is located in the Jemez mountains of northern New Mexico, at an altitude 
of 2630 m. It has a field of view of $\sim$2 sr and a duty cycle greater 
than 90\%. The effective area of Milagro is a function of zenith angle and 
ranges from $\sim10$ m$^2$ at 100 GeV to $\sim10^5$ m$^2$ at 10 TeV. A sparse 
array of 175 4000-l water tanks, each containing an individual PMT, was added in 
2002. These additional detectors, known as ``outriggers,'' extend the physical 
area of Milagro to 40000 m$^2$, substantially increasing the sensitivity of the 
instrument and lowering the energy threshold. The angular resolution is approximately 
0.75$^\circ$.

\section{Recent Results}

\subsection{Diffuse VHE emission from the Galactic Plane}

\begin{figure*}[t]
  \begin{center}
    \begin{tabular}{ll}
\includegraphics[width=60mm]{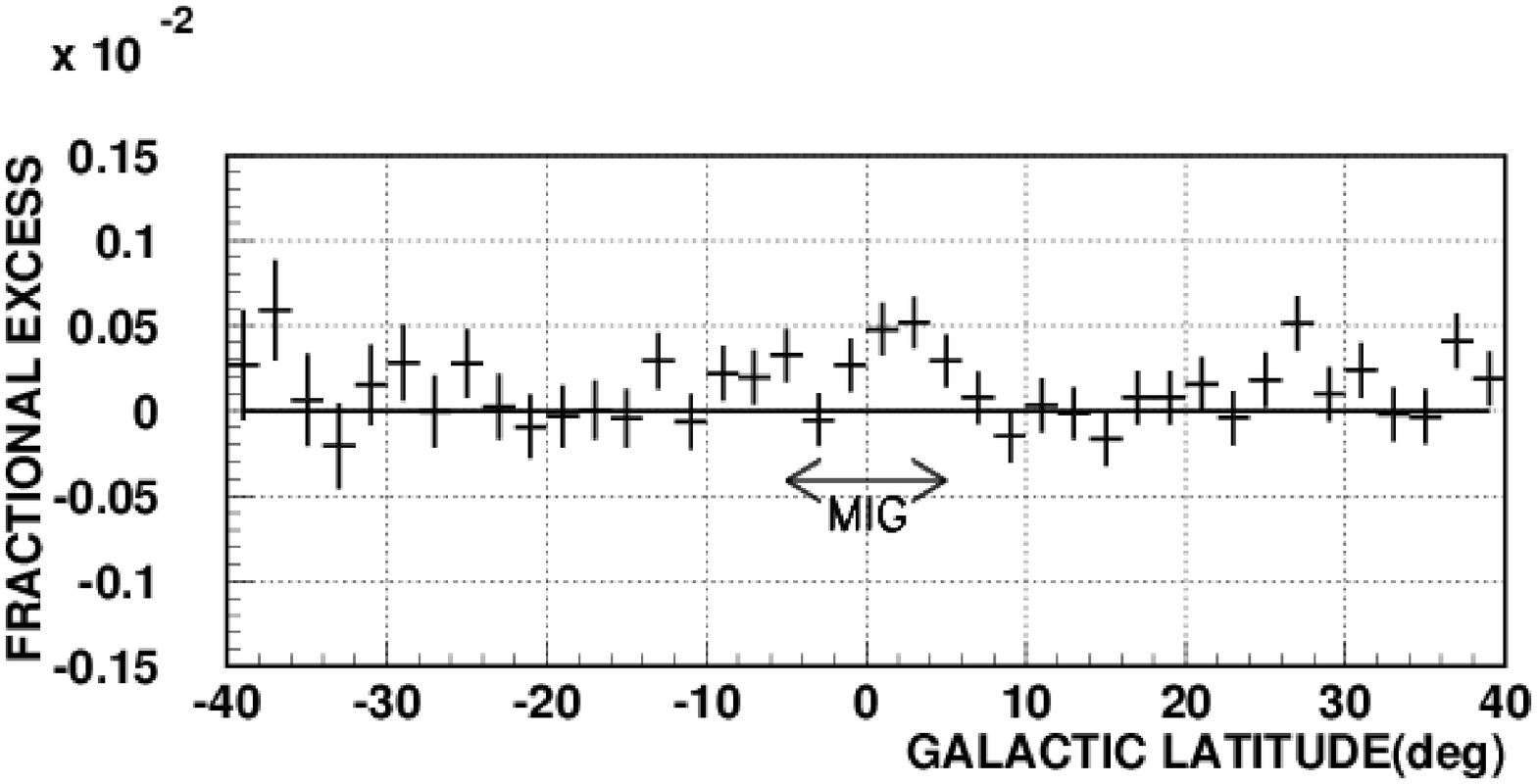} &
\includegraphics[width=60mm]{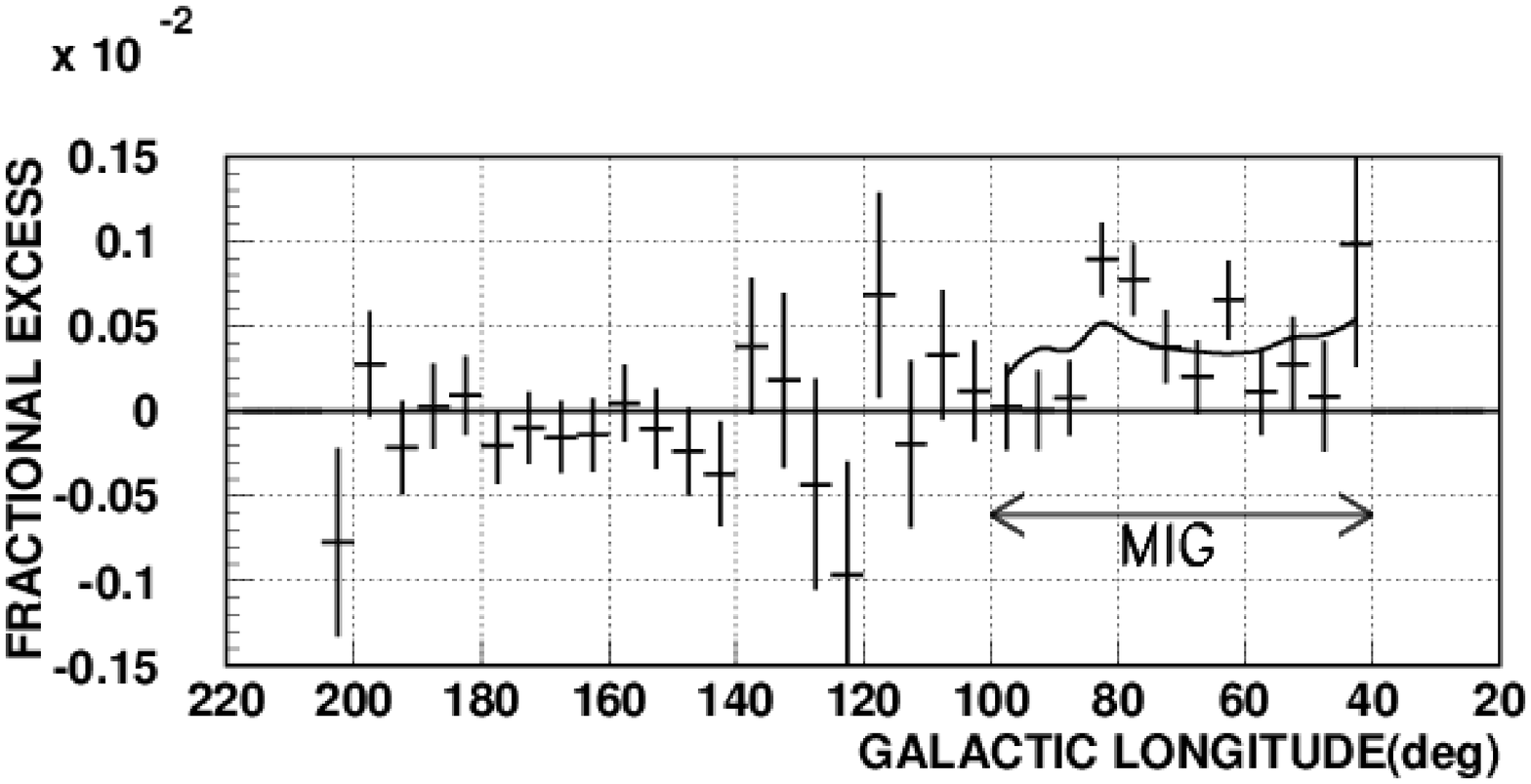} \\
    \end{tabular}
\caption{
{\bf Left} -- Profile of the fractional excess as a function of 
Galactic latitude (for Galactic longitude 40$^\circ$ to 100$^\circ$). 
{\bf Right} -- Profile of the fractional excess as a function of Galactic 
longitude (for Galactic latitude -5$^\circ$ to 5$^\circ$). The EGRET 
longitudinal source shape is superposed. (Figs. from Atkins et al. 2005)
}
\label{profile}
\end{center}
\end{figure*}

Diffuse emission from the Galactic plane is the dominant source in the gamma-ray 
sky.\cite{hunter97} Most of the diffuse VHE emission from the Galactic plane 
is thought to be produced by the interaction of cosmic-ray hadrons with interstellar 
matter. The flux measured by EGRET below 1 GeV fits models well, but that measured 
between 1 and 40 GeV is significantly larger than what is predicted by most models. One 
possible explanation for this enhanced emission is the inverse-Compton scattering of 
cosmic-ray electrons.\cite{1997JPhG...23.1765P} If this turns out to be the dominant 
source of diffuse gamma-ray emission from the Galactic plane, then the flux at TeV 
energies could be an order of magnitude higher than previously thought. 
Using 36 months of data, from 19 July 2000 to 18 July 2003, we looked at the inner 
(40--100$^\circ$) and outer (140--220$^\circ$) regions of the Galaxy. While the outer Galaxy 
shows no significant excess, the inner Galaxy shows a 5$\sigma$ (5 standard deviations). 
excess.\cite{2005AIPC..745..269S,atkins05} Fig.~\ref{profile} shows the profile in 
latitude for the longitude band (40--100$^\circ$) of the inner Galactic region 
(left panel) and the profile in longitude for the latitude band (-5$^\circ$ to 5$^\circ$) 
of the inner Galactic region (right panel), where the enhancement can be seen just north 
of the equator. The region of the inner Galaxy shows an enhancement along and just north of 
the Galactic equator. This is the same region where EGRET detected the strongest signal in 
the 100 MeV energy range. The 5$\sigma$ excess is seen by summing the entire inner Galaxy 
with a +/- 5$^\circ$ latitude band, as suggested by the EGRET results. The 
Milagro observation of the Galactic plane remains significant even when the region around 
the Cygnus Arm is excluded. This constitutes the first detection of the Galactic plane at 
TeV energies.

\subsection{Other extended sources}

A search for extended emission was carried out for the Milagro data collected between 
17 August 2000 and 5 May 2004.\cite{2005AIPC..745..657S} A set of standard cuts has been 
validated by observations of the Crab\cite{atkins03} and Mkn 421.\cite{atkins04} The 
background is computed from data collected at the same local detector coordinates, but at 
a different time, ensuring the celestial angles of the background event sample do not 
overlap with the source position under consideration. The method of Li \& Ma\cite{lima} is 
used to compute the significance of each excess. While the optimal square bin for detection 
of point sources with Milagro is 2.1$^\circ$ on each side,\cite{atkins04} to look for diffuse 
sources, the standard Milagro sky maps were searched using a range of bin sizes from 
2.1$^\circ$ to 5.9$^\circ$ in steps of 0.2$^\circ$. 20 separate searches were performed 
on the same maps, though the results are highly correlated. Monte Carlo simulations were 
used to compute the post-trials probability for each source candidate.

\begin{figure*}[t]
  \begin{center}
    \begin{tabular}{cc}
\includegraphics[width=40mm]{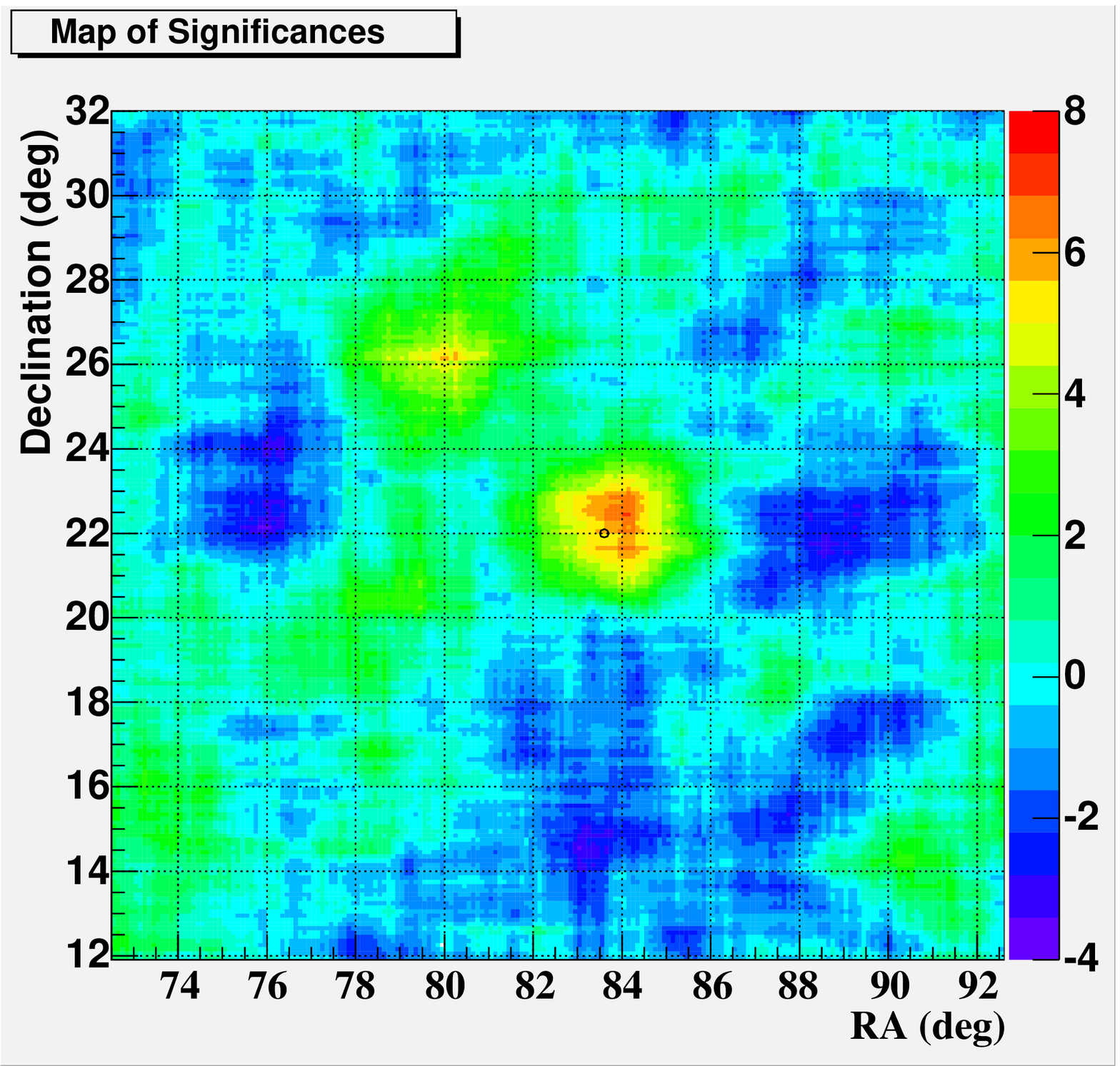} &
\includegraphics[width=40mm]{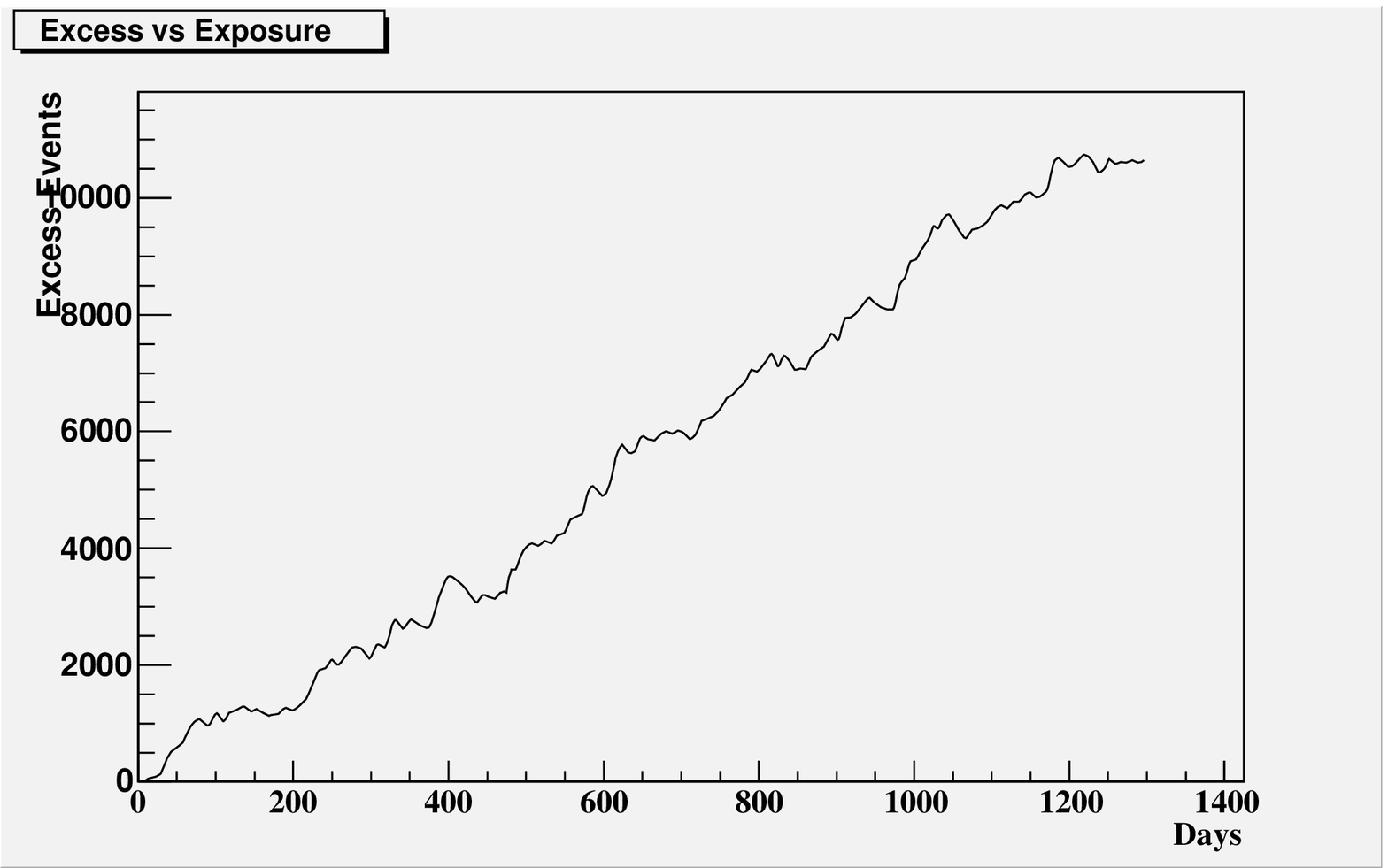} \\
    \end{tabular}
\caption{{\bf Left} -- Milagro significance map, with the Crab in the center and a TeV 
source coincident with 3EGJ0520+2556 to the left. {\bf Right} -- Cumulative 
excess as a function of time (Figs. from Smith et al. 2004)}
\label{egret_both}
\end{center}
\end{figure*}

{\bf 3EG J0520+2556}\quad The most significant candidate found in our search had a pre-trials 
significance of 5.9$\sigma$, located at RA=79.8$^\circ$ and 
Dec=26.0$^\circ$ and was identified using a 2.9$^\circ$ bin size. The probability of 
observing an excess this significant at any point in the sky at any bin size is 0.8\%. 
Fig.~\ref{egret_both} (left panel) shows the map of significances around the source, which 
is located $\sim$5.5$^\circ$ from the Crab. This candidate was first reported in 
2002.\cite{sinnis02} The cumulative significance using only data since it was first 
reported is 3.7$\sigma$. The right panel of Fig.~\ref{egret_both} shows the accumulation of 
excess events with time, indicating that the source shows no periods of significant flaring. 
This candidate is coincident with the EGRET unidentified source 3EG J0520+2556.

{\bf The Cygnus Arm}\quad The second extended source candidate is coincident with the region 
known as the Cygnus Arm, a spiral arm within our Galaxy that extends radially away from Earth. 
This is a dense region of gas and dust which was observed by EGRET as the 
brightest source of GeV gamma rays in the northern sky, with a 
diffuse GeV emission comparable to the Galactic bulge. Like in the Galactic 
plane region, VHE emission from the Cygnus Arm is thought to originate mainly from 
interactions of cosmic rays with the interstellar gas and dust. A 5.5$\sigma$ excess was 
detected using a 5.9$^\circ$ bin, at RA=308$^\circ$ and Dec=42$^\circ$. The probability of 
observing an excess this significant at any point in the sky at any bin size is 2.0\%. 
The excess observed with Milagro is inconsistent with a point source, and the number of 
excess events within the 5.9$^{\circ}$ bin corresponds to approximately twice 
the Crab flux. Like the case of 3EG J0520+2556, the accumulation of the excess is 
steady, and no evidence for flaring is observed. While this is an extremely bright 
region, making it the hottest spot in the Galactic plane, it is not surprising that it 
has not been detected yet by any of the Atmospheric Cerenkov Telescopes (ACTs), given the 
diffuse nature of the source and the limited field of view of such telescopes.

\subsection{Gamma Ray Bursts}

Many GRB models predict a fluence at TeV comparable to that at MeV 
scales.\cite{dermer00,pilla98,zhang01} Almost all GRBs are detected between 20 keV and 
1 MeV, though several have been observed above 100 MeV by EGRET, indicating that their 
spectrum extends at least out to 1 GeV.\cite{dingus01} Recently, a second component was 
found in one burst\cite{gonzalez03} extending out to at least 200 MeV and with a much 
slower temporal decay than the main burst. It is unclear how high in energy this component 
extends and whether it is similar to the inverse Compton peak seen in TeV sources. Above 
100 GeV, no conclusive emission has been detected for any single GRB. A 
search for counterparts to 54 BATSE bursts with Milagrito, a prototype of Milagro, found 
evidence for emission from one burst, with a significance slightly greater 
than 3 $\sigma$.\cite{atkins00} Twenty-five satellite-triggered GRBs occurred within the 
field of view of Milagro between January 2000 and December 2001. Due to the absorption of 
high-energy gamma rays by the extragalactic background light,\cite{stecker98,primack99} 
detections are only expected to be possible for redshifts less than $\sim$0.5. No significant 
emission was detected from any of these bursts.\cite{atkins05b} Between January 2002 and 
December 2004 only 10 well-localized GRBs were within 45$^\circ$ of zenith at 
Milagro, due to the demise of BATSE.\cite{2005AIPC..745..597S}. However, in the five months 
since the launch of the Swift satellite, there have been an additional 10 well-localized 
bursts, several of them with redshift information\footnote{A great resource for GRB 
localizations is J. Greiner's web page http://www.mpe.mpg.de/~jcg/grbgen.html}. Analysis 
of the most recent bursts is still under way but this much larger sample of bursts 
provided by Swift, especially those with a measured redshift, will allow us to either 
conclusively detect emission from bursts or place powerful constraints on 
the VHE emission from these bursts.

\section{Future}

Milagro has pioneered the water Cherenkov technique for the detection
of extensive air showers and has demonstrated the advantages that such a technique
has over traditional ACTs (e.g. higher duty cycle and much larger field of view). The 
HAWC (High Altitude Water Cherenkov)\cite{sinnis04,2005AIPC..745..234S} array is 
the next generation all-sky VHE gamma-ray telescope. This future detector would be located 
at an extreme altitude ($>$4km asl) and have a large area ($\sim$40,000 m$^2$), making it 
able to detect GRBs at redshifts $>$1, observe flares from active 
galaxies as short as 15 minutes in duration, and survey the overhead sky at a level 
of $\sim$ 30 mCrab in a year.
\\
\\
%\section*{Acknowledgments}
{\bf Acknowledgments}\quad
Many people helped bring Milagro to fruition.  In particular, we
acknowledge the efforts of Scott DeLay, Neil Thompson and Michael Schneider. 
This work has been supported by the National Science Foundation (under grants 
PHY-0075326, %Milagro Operations
-0096256, %UW-Madison
-0097315, %LANL via UMD 
-0206656, %NYU current; previous is PHY-9901496
-0245143, %UCSC; previous is PHY-0070927
-0245234, %UCI; previous is PHY-0070933
-0302000, %UMD
and
ATM-0002744) %UNH
the US Department of Energy (Office of High-Energy Physics and 
Office of Nuclear Physics), Los Alamos National Laboratory, the University of
California, and the Institute of Geophysics and Planetary Physics.

\end{document}